# Patient-centric HetNets Powered by Machine Learning and Big Data Analytics for 6G Networks

Mohammed S. Hadi, Graduate Student Member, IEEE, Ahmed Q. Lawey, Taisir E. H. El-Gorashi, and Jaafar M. H. Elmirghani, Senior Member, IEEE

School of Electronic and Electrical Engineering, Institute of Communication and Power Networks, University of Leeds, Leeds LS2 9JT, U.K

Corresponding author: Mohammed S. Hadi (elmsha@leeds.ac.uk).

This work was supported in part by the Engineering and Physical Sciences Research Council (EPSRC), in part by the INTERNET under Grant EP/H040536/1, in part by the STAR Projects under Grant EP/K016873/1 and in part by the TOWS Projects under Grant EP/S016570/1.

**ABSTRACT** Having a cognitive and self-optimizing network that proactively adapts not only to channel conditions, but also according to its users' needs can be one of the highest forthcoming priorities of future 6G Heterogeneous Networks (HetNets). In this paper, we introduce an interdisciplinary approach linking the concepts of e-healthcare, priority, big data analytics (BDA) and radio resource optimization in a multi-tier 5G network. We employ three machine learning (ML) algorithms, namely, naïve Bayesian (NB) classifier, logistic regression (LR), and decision tree (DT), working as an ensemble system to analyze historical medical records of stroke out-patients (OPs) and readings from body-attached internet-of-things (IoT) sensors to predict the likelihood of an imminent stroke. We convert the stroke likelihood into a risk factor functioning as a priority in a mixed integer linear programming (MILP) optimization model. Hence, the task is to optimally allocate physical resource blocks (PRBs) to HetNet users while prioritizing OPs by granting them high gain PRBs according to the severity of their medical state. Thus, empowering the OPs to send their critical data to their healthcare provider with minimized delay. To that end, two optimization approaches are proposed, a weighted sum rate maximization (WSRMax) approach and a proportional fairness (PF) approach. The proposed approaches increased the OPs' average signal to interference plus noise (SINR) by 57% and 95%, respectively. The WSRMax approach increased the system's total SINR to a level higher than that of the PF approach, nevertheless, the PF approach yielded higher SINRs for the OPs, better fairness and a lower margin of error.

## I. INTRODUCTION

Brain strokes are one of the rising health issues and though they might cause significant disabilities to the patient, immediate treatment can effectively increase recovery chances [1]. According to statistics from England, Wales and Northern Ireland for 2016-2017, one-third of stroke patients arrived at the hospital unaware of the date and time their symptoms began. The severity of this matter is even starker when knowing that the average waiting time for a patient from the start of symptoms until hospital admission is 7.5 hours, with an additional 55 minutes for door-to-needle time (the time between arriving at an emergency department and having an anesthetic administered). Adding to all that, the patient is loses 1.9 million neurons each minute until the treatment begins [2]. Thus, a proactive and timely diagnosis is vital. Big data analytics (BDA) and machine learning (ML) methods can be optimally utilized to process disparate data such as patient's electronic health record (EHR), diet, genetic data and their daily routine, and produce a quick and accurate diagnosis supporting medical personnel [3]. Thus, saving lives, improving the level of care, and lowering costs. An example of using BDA in healthcare is reported by the Medical Centre at Columbia University, where BDA is used to diagnose complications suffered from bleeding stroke caused by ruptured brain aneurysms. Using physiological data, complications were indicated 48 hours in advance, giving the medical staff sufficient time to treat the complications [4]. Given its importance in saving people's lives, the last five years witnessed a growing interest in the subject of using ML algorithms for stroke prediction as illustrated in Fig. 1.

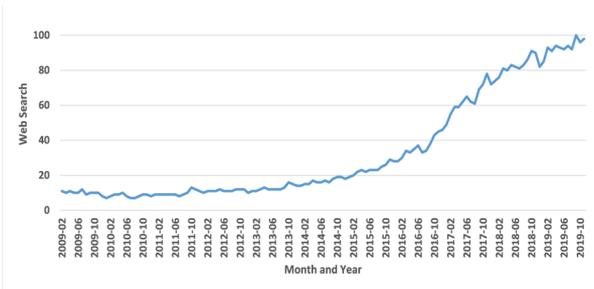

**FIGURE 1. Google trends for stroke prediction in the field of machine learning**

The topic of remote out-patient (OP) monitoring is largely dependent on two factors; precise medical Internet of Things (IoT) sensors and a reliable network connection to convey the relevant data. A cellular connection is preferred over Wi-Fi or wired connections as it does not restrict the patient's mobility. Nevertheless, cellular connections can experience channel fading and path loss where the connection can become unreliable or cannot be established due to a very low signal to interference plus noise ratio (SINR). A slow fading channel may indicate that the signal level is inadequate at the instance(s) when an OP's critical data must be conveyed urgently to the healthcare provider.

In this work, we are proposing to use the OP's data to serve a dual purpose. In addition to diagnostics, it would guide the network operator to the OPs with the most pressing needs. Hence, radio resources can be allocated to them. We contend that maintaining a high-quality connection between the OP's medical IoT and the medical provider is a step towards transforming conventional heterogeneous networks (HetNets) into a cognitively-personalized e-healthcare-centric service. Building self-adaptive, intelligent, and self-aware network is an operator's high-level objective. Therefore, ML can endow the network the capability of learning from experience and improving its performance. Therefore, BDA can transform the network from being reactive to predictive [5]. Topics that discusses patient monitoring, radio resource allocation, prioritization, fairness, and ensemble-aided disease risk prediction are popular in the literature across several disciplines. However, proposing a HetNet optimization framework that incorporates all the above is, to the extent of our knowledge, unique.

The objective of both approaches presented in this work is to maximize the system's overall SINR, both of which are governed by a number of power and physical resource block (PRB) assignment constraints. Nevertheless, OP prioritization is implemented in the assignment process by allocating the OPs with PRBs of gains relative to their medical state (i.e., the stroke likelihood) acquired from an ensemble system. The main contributions of this paper are: (i) extending our previous work in [6] to include a larger dataset, and the incorporation of new ML algorithms including decision tree (DT), logistic regression (LR), and the ML algorithm we deployed in [6] (naïve Bayesian (NB) classifier) in an ensemble system where a soft voting (SV) classifier resides; (ii) rigorously scrutinizing the classifiers' performance by conducting various tests of accuracy, recall, specificity, false-positive rate, false-negative rate, negative prediction rate, precision, and F1 score. Furthermore, reporting the cross-validation test scores for all datasets; (iii) extending the work in [7] to study the effects of inter-cell interference in HetNets, where we also added a reliability-aware aspect to the PF approach; (iv) testing the fairness among users, and conducting the required sensitivity analysis over 300 instances. The paper is organized as follows. Section II explores the related work. Section III illustrates the system model, the ML algorithms, and the problem formulation. Section IV presents and discusses the results. The paper concludes with Section V.

## II. Related Work

Our proposed system is of an interdisciplinary nature spanning the use of machine learning and big data analytics in cellular network design. In particular it is concerned with the use of machine learning and big data analytics in healthcare, particularly, disease prediction and hence applying the knowledge gained in the design of cellular wireless system. Therefore, we investigate both parts in this section. This section concludes with a third part highlighting the proposed link between the former two parts where we infused the two subjects by proposing a system that optimizes the uplink of a HetNet considering the intrinsic needs of the human users, in this case, their medical state.

### A. *Using machine learning/big data analytics in network design/optimization*

Significant effort is dedicated currently to endowing wireless networks with the ability to seamlessly prioritize cellular users to serve them accordingly. Utilizing big data analytics for network design was thoroughly discussed in our survey paper where we observed that the wireless field received the highest level of attention in this area [8]. Optimizing the resource allocation within the network can focus on several areas, such as spectrum allocation optimization, beamformer design, power optimization, backhaul management, computing resource optimization, and cache optimization. In this paper, we will concentrate on the spectrum allocation due to its relevance to our present work. The authors in [9] employed a genetic algorithm (GA) and support vector machines (SVM) and proposed a network planning tool. They proposed a metric which is the number of physical resource block per Megabit (i.e., PRB/Mb) and their target was to maintain the quality of service (QoS) while minimizing that metric. Having a faster dynamic spectrum allocation decision in a cloud-based radio access network (C-RAN) was the goal of the authors in [10]. They proposed the use of regression analysis applied to big data collected from a monitoring system at Sofia Airport to predict the spectrum occupancy and usage activity in predefined frequency bands. Optimizing the spectrum allocation, peer discovery, and route selection from a delay perspective was the goal of the authors in [11]. They predicted the vehicle trajectories using interacting multiple model (IMM) estimation and a multi-Kalman filter (MKF) operating on big data generated by geographic positioning system (GPS) and geographic information system (GIS). Game theory is used in the optimization part, where a coalition formation game is formulated. The authors in [12] proposed a scheme to solve the coexistence problem of the Wi-Fi and LTE unlicensed (LTE-U) in the unlicensed spectrum. The scheme uses Q-learning (QL) to dynamically allocate blank subframes while maintaining the frame size (i.e., reducing the subframe length). Further, the authors proposed to share the transmission-related information to let the LTE-U decide when to allocate the blank subframes, and to have the blank subframe number dynamically adjustable relative to the Wi-Fi traffic size.

### B. *Using machine learning and big data analytics in healthcare and stroke prediction*

Cardiovascular disease (CVD) prediction using ML/BDA techniques has been comprehensively discussed in prior literature. The authors of [12] proposed an approach to help patients with Parkinson's disease. Given the fact that the loss of flexibility is a sign of disease progression, the authors proposed a system that analyzes big data collected from body and 3D sensors (e.g., Microsoft Kinect). This gives healthcare providers an opportunity to track treatment effectiveness and disease development in real-time. The authors of a comprehensive study in [13] employed a NB classifier to predict heart disease problems. The same classifier was corroborated by a number of cardiologists in [14] and the classifier's accuracy had an agreement of more than 80% with the health outcomes of the respondents. K-nearest neighbor (KNN), DT, artificial neural networks (ANN), and SVM algorithms were compared in [15] for the detection of small ischemic stroke in brain computerized tomography (CT) images, where SVM reported the highest prediction accuracy. More than 800 subjects participated in the research in [16] where a dataset comprising 50 risk factors was collected. KNN and C4.5 DT algorithms were used to analyze the data using a WEKA platform to predict stroke incidence. The C4.5 DT algorithm returned the highest accuracy. The authors in [17] employed naïve Bayes, ANN, and DT methods and the DT reported the highest accuracy score. A proof-of-concept study was conducted in [18] to predict the outcome of stroke thrombolysis using SVM . It analyzed a sample of 116 brain CTs and clinical records and compared the results with prognostication tools. The results were in favor of the SVM. Predicting the intensive care unit (ICU) transfer decision for stroke in-patients was the target for the researchers in [19]. DT, SVM, ANN, and logistic regression (LR) ensemble were employed for that purpose. The results were then compared to a generalized boosted model (GBM). The results showed the DT to have a higher accuracy than the other models. However, the GBM ensemble showed a better result than the DT. To classify the type of stroke, the authors of [20] proposed to utilize case sheets for patients' symptoms using data mining methods, namely, tagging and maximum entropy methodologies and machine learning to classify the stroke type. Towards that end, the authors used SVM, boosting and bagging, ANN, and random forests algorithms. The results showed that the ANN yielded higher classification accuracy than the rest of the algorithms.

### C. *Bridging the gap between the two topics*

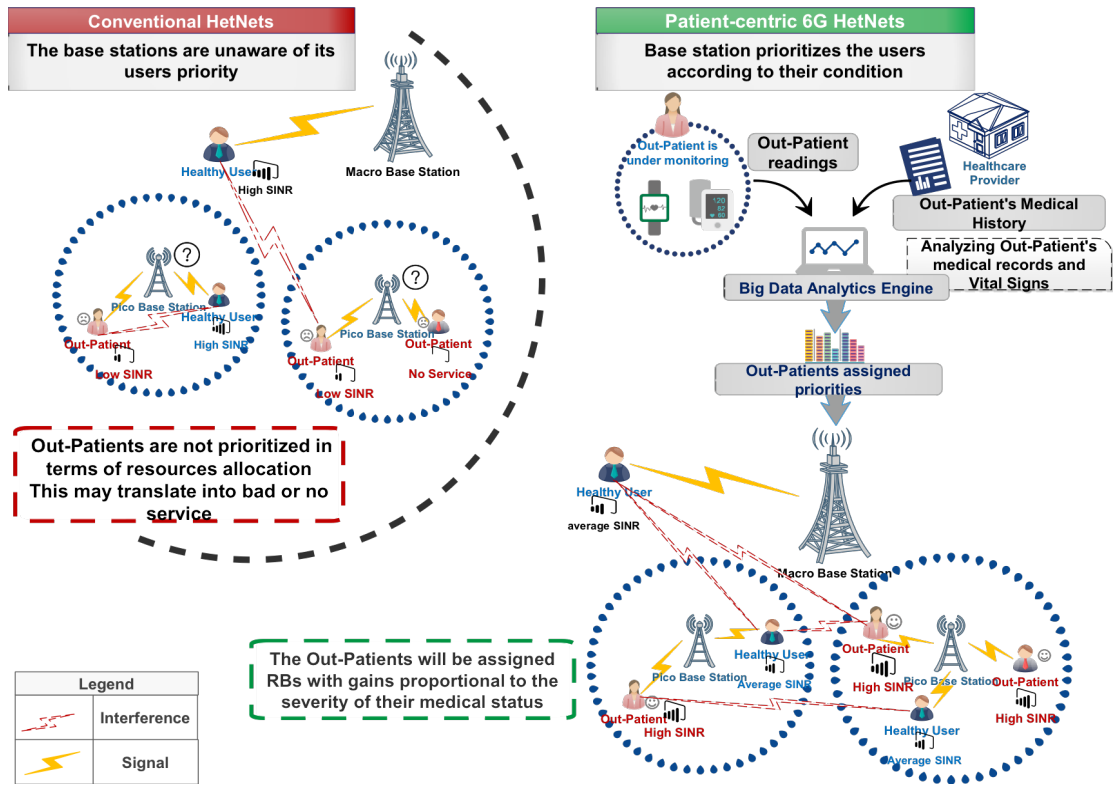

FIGURE 1. Patient-Aware 6G HetNet

Despite the highlighted surge in the literature on self-optimized and ML-assisted network optimization, this prior research is still heavily reliant on either terminal-level or network-level features. Furthermore, the works presented are mostly agnostic to the users and their needs. A recent work presented by the authors in [21] proposed a human brain-aware optimization approach to allocate radio resources. They developed a probability distribution identification (PDI) learning method to predict the delay perceived by human users and quantify the reliability of this prediction. Based on a proposed closed-form expression linking wireless physical layer metrics and reliability measure, and by using a Lyapunov-based brain-aware optimization approach the authors were able to allocate radio resources to human users according to their delay perception. In a previous work, we proposed in [6] the use of readings of blood pressure (systolic and diastolic), total cholesterol levels and daily smoking rate acquired by non-invasive medical IoT sensors and train a NB classifier on the OP's medical record to predict the likelihood of a stroke. This likelihood is then feedback as a priority factor in a radio resource optimization model for an LTE-A network guaranteeing the assignment of high-quality PRBs to the OPs. Furthermore, we extended the previous work to include multi-tier HetNets in [7] with a spectrum partitioning strategy [22] so that inter-tier interference is mitigated. Moreover, the system response was investigated over seven different current states resulting in different priority levels granted to the OPs.

## III. OP-Centric Network Optimization Framework

In this section, we introduce the system model, before explaining the role of the machine learning algorithms and all of the stages of data preparation needed before the data can be usable in the proposed model. The problem formulation concludes this section, where we present the main mathematical equations in terms of objective functions and constraints.

### A. *System Model*

In this work, we consider a scenario of a HetNet consisting of a macro base station (MBS) and two neighboring pico base stations (PBS) operating in an urban environment. The MBS coverage range is a radius of 300-600 meters whilst the PBS has a coverage radius of 40-100 meters. In a previous work in [7], we assumed the adoption of a spectrum partitioning strategy [22] to mitigate the impact of inter-tier interference on the PBS users caused by the MBS users. In this work, we consider the effects of the inter-tier interference. The users belong to two categories: healthy (normal) users, and OPs as illustrated in Fig. 2 which shows a 6G HetNet scenario with BDA. As in a real-life scenario, the users are randomly scattered around the base stations at different distances resulting in different received power levels at the base station from connected UEs. If an OP is assigned a low-level SINR channel, the healthcare provider may not be notified in an emergency and the response is thus delayed. Here, a patient suffering a stroke loses 1.9 million neurons per minute before the treatment starts [2]. Therefore, the objective is to assign high-gain PRBs to the OPs according to the severity of their medical status (i.e., stroke likelihood). The latter is computed in a cloud-based BDA engine according to the procedure shown in Fig. 3. Thus, OPs that are prioritized over normal users will have higher spectral efficiency due to their high SINR values. This, in turn, will achieve higher throughput (since spectral efficiency is directly proportional to throughput). Hence, the OPs will be able to send their data with minimal delay.

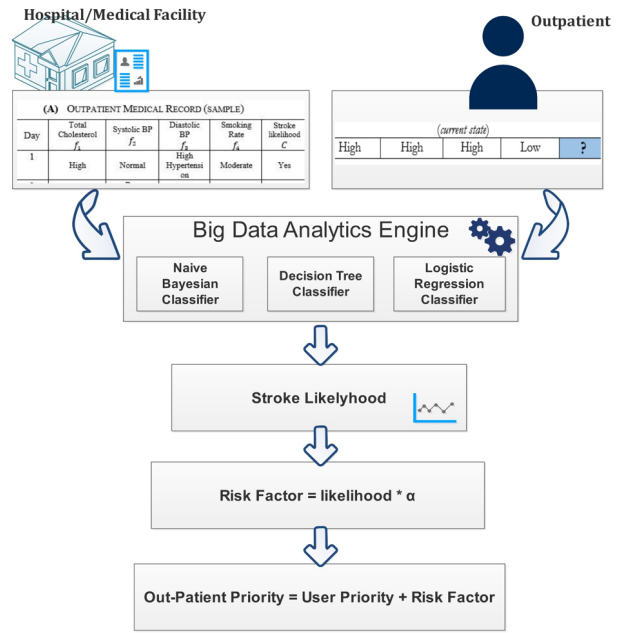

**FIGURE 2. Out-Patient Priority Calculation Procedure**

### B. *Machine learning algorithms*

In this work, we use an ensemble system comprising three supervised learning classifiers, namely, a NB classifier, a DT classifier, and a LR classifier that work on the OP's dataset and feed their predicted probabilities of stroke to a soft voting (SV) classifier. Given a certain feature vector (representing the OP's current state), each of the aforesaid classifiers yields a probability of stroke. Using ensemble learning, those classifiers can be combined into a single predictive model with higher accuracy, and thus, higher confidence is achieved in the predicted results.

Table I (A) shows a sample of the dataset (i.e., medical record) of a single OP. Feature variables $f_1, ... f_4$ characterize the main contributors to a stroke stated in [23, 24] and they are (Blood Pressure (BP), Cholesterol Level, and Smoking Rate). The Framingham cardiovascular cohort study [25] is used to populate the dataset of the individual OPs. The Framingham study contains readings for more than 3000 persons. Due to regulatory and privacy reasons, it was not possible for us to acquire cohort medical records belonging to several *individual* patients. Therefore, we segmented parts of the Framingham dataset in [26] to represent several OPs.

**TABLE I.** OUTPATIENT MEDICAL RECORD (SAMPLE)

| Day | Total Cholesterol $f_1$ | Systolic BP $f_2$ | Diastolic BP $f_3$ | Smoking Rate $f_4$ | Stroke $C$ |
|---|---|---|---|---|---|
| 1 | High | Normal | High Hypertension | Moderate | Yes |
| 2 | Normal | Pre-hypertension | Low | Heavy | No |
| ⋮ | ⋮ | ⋮ | ⋮ | ⋮ | ⋮ |
| 200 | Optimal | High Hypertension | Pre-hypertension | Light | No |

**(A) (CURRENT STATE)**

| instance | $f_1$ | $f_2$ | $f_3$ | $f_4$ | $C$ |
|---|---|---|---|---|---|
| 1 | Normal | Pre-hypertension | Normal | Heavy | ? |

The ranges depicted in Table I (A) are based on the ones in Table II. To be as medically correct as possible, the discretized values of $f_1, \ldots f_3$ are in line with governmental health institutes or official organizations (e.g., the American National Institute of Health and the British Stroke Association) [27-29]. As for $f_4$, it uses the ranges in [30].

**TABLE II.** FEATURE VALUES AND THEIR CORRESPONDING LEVEL

| Feature | Range | Level |
|---|---|---|
| **Total cholesterol Level (mg/dl)** [27] | <200 | Optimal |
|  | 200-239 | Normal |
|  | 240+ | High |
| **Systolic BP (mmHg)** [28] [29] | <120 | Normal |
|  | 120-139 | Pre-hypertension |
|  | 140+ | High Hypertension |
| **Diastolic BP (mmHg)** [28] [29] | <80 | Normal |
|  | 80-89 | Pre-hypertension |
|  | 90+ | High Hypertension |
| **Smoking rate (Cig/Day)** [30] | 1 - 10 | Light |
|  | 11 - 19 | Moderate |
|  | 20+ | Heavy |

### 1) Naïve Bayesian Classifier

The NB classifier is a probabilistic statistical classifier which uses a number of independent feature variables $f_i$ (e.g. total Cholesterol and Blood pressure levels) obtained from a historical dataset (i.e., the OP's medical record) to determine the likelihood of an incident $c$ (i.e. a stroke) as shown in Fig. 3. The classifier is termed naïve because it assumes the feature variables are unrelated to each other [31]. This classifier is chosen for the following reasons; (i) it has a track record in disease risk prediction as in [14] and [32], (ii) its low complexity incurs less computational burden, (iii) it is an ideal choice for any two-class concept with nominal features [33], (iv) it has proven accuracy in Cardio Vascular Disease (CVD) prediction compared to other approaches [34, 35], (v) it does not require large training datasets [36].

The classifier's *posterior probability* is given as

$$P(C = c | F_i = f_i) = P(C = c) \prod_{i=1}^{n} P(F_i = f_i | C = c) \quad (1)$$

where $P(C = c)$ represents *the prior probability* of stroke, and the *likelihood* of $F$ given $C$ is given in (2)

$$P(F_i = f_i | C = c) = \frac{\sum_{i=1}^{n}(C = c \wedge F_i = f_i)}{\sum_{i=1}^{n}(C_i = C_i)} \quad (2)$$

where the term $\prod_{i=1}^{n} P(F_i = f_i | C = c)$ depicts the *joint probability*.

### 2) Logistic Regression Classifier

The main distinctions between the NB classifier and the LR classifier is that; (i) it is fast and can produce a large change in response to the feature vector, (ii) it allows for large discrimination (i.e., a change in one feature may cause large effect). However, this also means that it suffers from high sensitivity to feature vector values. This classifier is a popular tool in disease prediction as in [19, 20, 37]. The logistic model is based on the logistic function given in (3). This function is zero when $x$ is -∞, whereas the function is 1 when $x$ is +∞.

$$f(x) = \frac{1}{1 + e^{-x}} \quad (3)$$

This range is the primary reason for selecting the logistic model to estimate the probability. The index of combined features is $x$ and it is given as a linear sum as shown in (4).

$$x = \beta_0 + \beta_1 f_1 + \beta_2 f_2 + \ldots + \beta_n f_n \quad (4)$$

where $\beta_0$ represents the $y$ intercept and $\beta_1 \ldots \beta_n$ are the regression coefficients, $f_1, \ldots f_n$ depict the feature variables, and $n$ is the total number of features in the prediction model (in this work, $n = 4$) [38]. The conditional probability can be written as:

$$P(C = c | F_i = f_i) = \frac{1}{1 + e^{-(\beta_0 + \sum_{i=1}^{n} \beta_i f_i)}} \quad (5)$$

where $P(C = c | F_i = f_i)$ represents the conditional probability of a certain class variable $C = c$ given a feature vector $FV$. Therefore, if $C = 1$ then the conditional probability for $C = 0$ is $P(C = 0 | F_i = f_i) = 1 - P(C = 1 | F_i = f_i)$. The values of the line coefficients (i.e., $\beta_0 \ldots \beta_n$) cannot be solved analytically, therefore, we used solvers to navigate the search space.

### 3) Decision Trees Classifier

The DT construction procedure is done by splitting the dataset into descendant subsets. The splitting continues on repeated splits of the descendant subsets. The notion behind the tree methods is to have a set of partitions so that the best class can be determined. The partitions are performed so as to choose the splits in a way that guarantees that the leaves are *purer* than the parent node [39]. DT classifies vectors by sorting them, starting at the root of the tree down to some leaf nodes. In this tree, each node specifies a test of some input feature of the vector, and each branch descending from that node corresponds to one of the possible values for this feature [31]. The reasons for choosing DTs are; (i) their ability to implicitly perform feature selection or variable screening [20, 40, 41], (ii) they are uncomplicated to understand, interpret and, visualise, (iii) tree performance is not affected by nonlinear relationships between parameters, (iv) their track record in the stroke prediction literature as in [16, 17, 42].

The purity is measured using a *Gini index* which is used as an *attribute selection measure* where the ranking per attribute is given. The feature (attribute) with the best score is selected as the splitting feature for the given data subset. Splitting is done according to an impurity test conducted on a feature and a splitting subset (e.g., selecting two levels out of three $\{moderate, heavy\} \subset smoking$ or $\{moderate, heavy\} \subset V_{F_4}^r$ to be on a leaf while the remaining $\{low\} \subset V_{F_4}^r$ level is assigned to the other leaf). The binary split resulting in the maximum reduction in impurity (i.e., highest information gain) is selected as the splitting criterion. The *Gini* measure is given in (6).

$$Gini(\gamma) = 1 - \sum_{i=1}^{m}(p_i^\gamma)^2 \qquad (6)$$

where $p_i^\gamma$ is the probability of a feature vector in training dataset $\gamma$ belonging to class $C_i^\gamma$ of a total number of $m$ classes. The probability of an outcome of a certain class is given in (7) and the sum is calculated over $m$ classes [43], where

$$p_i^\gamma = \frac{|C_i^\gamma|}{|\gamma|}. \qquad (7)$$

It should be noted that the possible number of subsets is $2^{V_{F_i}^r} - 2$ (excluding the empty subset and the all $V_{F_i}^r$ subset), where $V_{F_i}^r$ represents the number of distinct values feature $F_i$ can have. However, in binary splits, this number is further reduced by omitting the cases where certain values are not included (e.g., assigning $\{moderate\} \subset V_{F_4}^r$ to one leaf and $\{heavy\} \subset V_{F_4}^r$ to another leaf and leaving the value $\{low\} \subset V_{F_4}^r$ unassigned. The weighted sum of the impurity is calculated for each resulting partition. Thus, if a feature $F_i$ partitions the dataset $\gamma$ into $\gamma_1$ and $\gamma_2$, then the Gini index of $\gamma$ is given in (8).

$$Gini_{f_i}(\gamma) = \frac{|\gamma_1|}{|\gamma|}Gini(\gamma_1) + \frac{|\gamma_2|}{|\gamma|}Gini(\gamma_2) \qquad (8)$$

The subset with the minimum impurity (i.e., Gini) for that feature is selected as its splitting subset. The same strategy is employed when using features with continuous values where each possible splitting point must be considered. Thus, extra computational resources will be required compared to the prior case.

The impurity reduction incurred by the binary split on feature $F_i$ is given in (9) is given by

$$\Delta Gini(f_i) = Gini(\gamma) - Gini_{F_i}(\gamma). \qquad (9)$$

After forming the DT for an outpatient, the probability of a given vector of medical measurements is evaluated by tracing the decisions down the tree till the leaf where this vector belongs is reached. The probability of a given leaf is then evaluated as.

$$P(C = c|F_i = f_i) = \frac{\Gamma_{z,C_i}}{\sum_{i=1}^{n}\Gamma_{z,C_i}} \qquad (10)$$

where $\Gamma_{z,C_i}$ denotes the number of samples in a leaf belonging to outpatient $z$ having class $C_i$. The denominator represents the total number of samples of all classes in a given leaf.

### 4) Ensemble model

Ensemble methods train multiple learners on the same dataset to classify the same feature vector(s). The original goal of using ensemble systems is comparable to the way a person seeks advice from several trusted individuals. Hence, this reinforces the confidence that the decision made was the right one. Similarly, an ensemble of classifiers can be employed to increase the classification accuracy. Ensemble systems provide a method to incorporate various opinions, sometimes weighing them differently before reaching a concluding verdict. Individual classifiers may have different errors, however, they generally agree in terms of their classification decision. Therefore, averaging the classifiers' outputs results in averaging the error component, consequently reducing the classification error [44, 45] and balancing out the individual weaknesses of equally well-performing models [46]. The ensemble architecture of the soft voting (SV) classifier we employed in this work is illustrated in Fig. 4. The NBC, LR, and DT serve as base classifiers and their probabilities are then averaged to produce the voted probability denoted by $P_{voting}$. To calculate this probability, let the probability yielded by each base classifier $CLF_i$ given in (1), (5), and (10) be $P_{CLF_1}$, $P_{CLF_2}$ and $P_{CLF_3}$, respectively. Since all base classifiers are treated evenly, the SV classifier calculates the probability as in (11).

$$P_{voting} = \frac{1}{|CLF_i|}\sum_{i=1}^{|CLF_i|} P_{CLF_i}(C = c|F_i = f_i) \qquad (11)$$

where $P_{voting}$ denotes the ensemble-calculated, averaged-conditional-probabilities.

In order to provide weights to the MILP so that the OPs are assigned higher gain PRBs, a base user priority $UP_k$ of 1 is assigned to normal users while OPs are assigned the base weight *plus* another weight derived from the multiplication of a weight parameter $\alpha$ by the voted stroke likelihood $P_{voting}$ thus, granting an effective-yet-reasonable priority:

$$UP_k = 1 + \alpha \cdot P_{voting}. \qquad (12)$$
$$\forall k \in \mathcal{K}: z = k, k \succ NU$$

The OP's *updated* priority is given in (12). Using different values of $\alpha$ impacts the system response accordingly in terms of the OPs' SINR levels as shown in the results section.

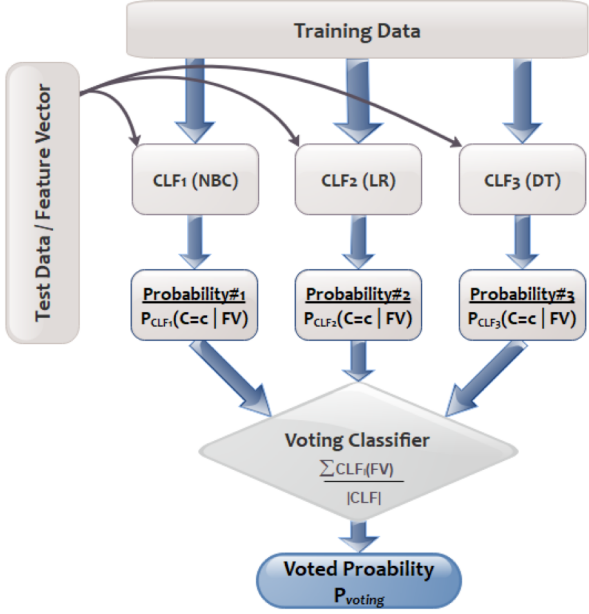

**FIGURE 3.** Ensemble Architecture

### C. *Problem Formulation and Model Parameters*

Using our track record in MILP optimization in [6, 7, 47-54], and physical layer modeling in [55-60], we developed the following MILP models introduced in [6] to optimize the cellular system resource allocation for OPs and normal users. We consider a scenario where the OPs monitoring system operates in a HetNet supported by $B$ base stations denoted by set $\mathcal{B} = \{1, ..., B\}$ including both MBS and PBS types, operating at channels with 1.4 MHz bandwidth. Each base station $b$ has $N$ PRBs depicted by set $\mathcal{N} = \{1, ..., N\}$. The network serves a total of $K$ users (normal and OPs) denoted by set $\mathcal{K} = \{1, ..., K\}$ by allocating PRB $n$ to connect to BS $b$ in an instant in time. The goal is to optimize the uplink of the HetNet, so that the OPs are prioritized over healthy users; hence, allocating them high-gain PRBs.

We formulate this problem as a MILP model. Table III defines the sets, parameters, and variables used in the network optimization problem formulation.

**TABLE III.** SYSTEM SETS, PARAMETERS, AND VARIABLES

| Sets | |
|---|---|
| $\mathcal{K}$ | Set of users. |
| $\mathcal{N}$ | Set of physical resource blocks. |
| $\mathcal{B}$ | Set of base stations. |
| $\mathcal{F}$ | Set of features in the learning dataset. |
| $C$ | Set of classes in the learning dataset. |
| $\mathcal{Z}$ | Set of outpatient users, $(\mathcal{Z} \subset \mathcal{K})$. |
| $CLF_i$ | Set of base classifiers |
| $V_{F_i}^r$ | Set of values feature $F_i$ can take in the learning dataset. |
| **Parameters** | |
| $CS_i$ | The current state of the patient in feature $i$ (e.g. Cholesterol value). |
| $UP_k$ | User priority ($UP_k$ =1 for normal users whereas $UP_k > 1$ is granted for OPs depending on their risk factor). |
| $Q_{k,n}^b$ | Power received from user $k$ using physical resource block $n$ at base station $b$. |
| $H_{k,n}^b$ | Rayleigh fading with zero mean and a standard deviation equal to 1 experienced by user $k$ using physical resource block $n$ at base station $b$. |
| $A_k^b$ | Signal attenuation experienced by user $k$ connected to base station $b$. |
| $PM$ | Maximum power allowed per uplink connection. |
| $P$ | Power consumed to utilize physical resource block $n$ to connect user $k$ to base station $b$. |
| $\sigma_{k,n}^b$ | Additive White Gaussian Noise (AWGN) power in watts experienced by user $k$ using physical resource block $n$ at base station $b$. |
| $P_{voting}$ | The probability of stroke calculated at the voting classifier. |
| $\alpha$ | Tuning factor. |
| $NU$ | The total number of normal users. |
| $\psi$ | The minimum SINR level. |
| **Variables** | |
| $X_{k,n}^b$ | Binary decision variable $X_{k,n}^b = 1$ if user $k$ is assigned physical resource block $n$ in base station $b$, otherwise $X_{k,n}^b = 0$. |
| $T_{k,n}^b$ | The SINR of user $k$ utilizing physical resource block $n$ at base station $b$. |
| $\phi_{m,n,k}^{w,b}$ | Non-negative linearization variable where $\phi_{m,n,k}^{w,b} = T_{k,n}^b X_{m,n}^w$. |
| $S_k$ | SINR of user $k$. |
| $L_k$ | Logarithmic SINR of user $k$. |

The user's uplink SINR of an OFDMA network can be expressed as in [61].

$$T_{k,n}^b = \frac{Q_{k,n}^b X_{k,n}^b}{\sum_{\substack{w \in \mathcal{B} \\ w \neq b}} \sum_{\substack{m \in \mathcal{K} \\ m \neq k}} Q_{m,n}^b X_{m,n}^w + \sigma_{k,n}^b} \quad (13)$$

Examining the numerator (i.e. signal), $Q_{k,n}^b X_{k,n}^b$ signifies the signal power received at the base station from user $k$. $X_{k,n}^b$ is a binary decision variable, $X_{k,n}^b = 1$ denotes the connection of user $k$ to PRB $n$ in base station $b$. The power received at base station $b$ from the interfering user(s) $m, m \neq k$, on the same PRB is $Q_{m,n}^b X_{m,n}^w$; while $X_{m,n}^w$ indicates an interfering user(s) $m$ connected to another BS $w, w \neq b$ on PRB $n$. The AWGN is given as $\sigma_{k,n}^b$. Equation (13) is graphically illustrated in Fig. 5.

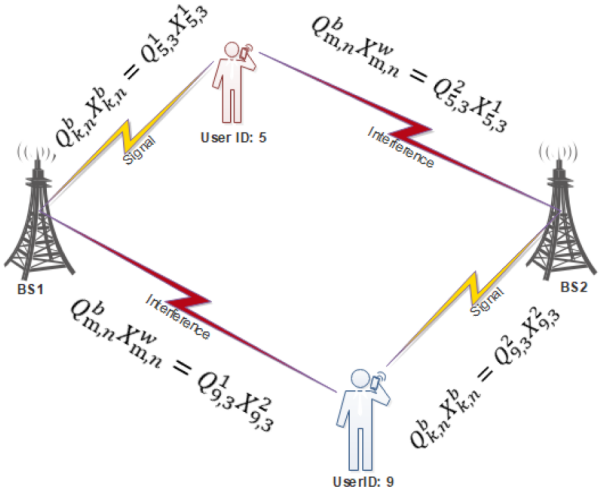

**FIGURE 4.** User Interference

Rewriting equation (13) gives a quadratic term resulting from the multiplication of two variables (Binary $X_{m,n}^w$ and Continuous $T_{k,n}^b$). Hence we follow [62] and define the variable $\phi_{m,n,k}^{w,b}$ that comprises all the indexes of both aforementioned (i.e., binary and float) variables as indicated in equation (14).

$$\sum_{\substack{w\in\mathcal{B}\\w\neq b}}\sum_{\substack{m\in\mathcal{K}\\m\neq k}} Q_{m,n}^b \phi_{m,n,k}^{w,b} + T_{k,n}^b \sigma_{k,n}^b = Q_{k,n}^b X_{k,n}^b \quad (14)$$

$$\forall\ k \in \mathcal{K}, n \in \mathcal{N}, b \in \mathcal{B}$$

where $\phi_{m,n,k}^{w,b} = T_{k,n}^b X_{m,n}^w$ and it represents the SINR of user $k$ with PRB $n$ connected to base station $b$ if there is an interfering user $m$ connected to the other base station $w$ with the same PRB $n$; otherwise, it is zero. Due to the introduction of $\phi_{m,n,k}^{w,b}$, it is imperative to linearize this term to solve the model using linear solvers such as CPLEX. Therefore, several linearization constraints are employed as given in [6].

We have developed three approaches to solve the resource allocation problem. The first approach, named WSRMax, uses an objective function that maximizes the Weighted Sum-Rate of the SINRs experienced by the users. The second approach implements fairness among cellular users by adopting a Proportionally Fair (PF) objective function. The third approach integrates the second approach with the concept of reliability. Thus, it is a reliability-aware PF where it sets an SINR threshold guaranteeing adequate QoS levels to normal users.

### 1) MILP Formulation for the WSRMax approach

In this approach, the objective is to maximize the system's overall SINR. This can be done by maximizing the SINRs of individual users.

#### a) Before Prioritizing the OPs

The OPs' risk factors introduced in the previous section are scaled into priorities (i.e. weights) and used to grant the OPs priority over other users. The MILP model is formulated as follows:

**Objective:** Maximize

$$\sum_{k\in\mathcal{K}}\sum_{n\in\mathcal{N}}\sum_{b\in\mathcal{B}} T_{k,n}^b\, UP_k \quad (15)$$

The objective in (15) targets the maximization of the weighted sum of the users' SINRs. The OPs have higher weights (i.e. priorities) than other healthy users and these weights are relative to the OPs' calculated risk factor.

Note that all the users share the same initial priority (i.e., $UP_k = 1$) as in (16). However, the OPs will have updated values according to their risk factor. This will ultimately drive the system into prioritizing the OPs over healthy users during PRB assignment. The mathematical formulations related to the OP weight (priority) calculation was illustrated in Section III.B.

$$UP_k = 1 \quad (16)$$
$$\forall\ k \in \mathcal{K}$$

**Constraints:**

The main constraints governing the MILP model's operation conform with our previous work in [6] and they ensure the following:

i- That the sum of powers per PRB connection $P$ for each user does not exceed the maximum allocated power per uplink connection $PM$.
ii- That each PRB in each BS is to be utilized by a maximum of one UE only, while each UE preserves the right to utilize more than one PRB when possible.
iii- Each user is allocated at least one PRB from any base station. Thus, no user is left without service. Furthermore, this stops the MILP from blocking interfering users to maximize the overall (network-wide) SINR.

#### b) After Prioritizing the OPs

In this approach, OPs' risk factors introduced in the previous section are scaled into weights to prioritize the OPs over other users. The MILP model is formulated in the same way as mentioned in the previous subsection. However, equation (12) is included in this model to represent the OPs' weights (i.e. priorities) while (16) is replaced by (17) to cover the normal users only, thus

$$UP_k = 1. \quad (17)$$
$$\forall\ k \in \mathcal{K}: 1 \leq k \leq NU$$

### 2) MILP Formulation for the PF approach

Maximizing the logarithmic sum of the user's SINRs is the objective in this approach. A slight decrease in the overall SINR might be observed (due to the nature of the natural logarithm) but with the added benefit of preserving fairness among normal users.

*a)    Before Prioritizing the OPs*

All users, in this case, are treated evenly, thus there is no prioritization in allocating the radio resources. However, keeping fairness among users still holds as a necessity. Since the only part that we are dealing with is the value of the individual user's SINR, and to simplify the manipulation of the equation before adding the natural logarithm, we introduce the optimization variable $S_k$, to serve as the SINR for each user $k$ as in [6], where the single-indexed variable $S_k$ substitutes the three-indexed variable $T_{k,n}^b$ as follows

$$L_k = \ln S_k.$$
$$\forall\ k \in \mathcal{K} \tag{18}$$

Calculating $L_k$ as a logarithmic function of the user's SINR $S_k$ is carried out in (18). Since the natural log is a concave function, and to maintain the linearity of our model, piecewise linearization was employed through a set of constraints as in [6].

The objective of this approach is given in (19):
**Objective**: Maximize

$$\sum_{k \in K} L_k \tag{19}$$

**Constraints:**

In addition to the previously-mentioned constraints, the PF approach employs a set of piecewise linearization relations implemented to linearize the concave function in equation (18). It should be noted that these relations resemble the line equation $y = mx + h$ where the line coefficients are selected as in [63]. It is worth noting that the number of constraints used in the linearization procedure is dictated by the total number of lines used to cover the linearized interval.

*b)    After Prioritizing the OPs*

The outpatients are prioritized in this case, and equation (18) is rewritten to reflect the change.

$$L_k = \ln S_k$$
$$\forall\ k \in \mathcal{K}: 1 \leq k \leq NU \tag{20}$$

Equation (20) shows that the log function is applied to normal users only. The OPs, on the other hand, are assigned weights instead.

**Objective:** Maximize

$$\sum_{k \in K, 1 \leq k \leq NU} L_k + \sum_{k \in K, k > NU} S_k UP_k \tag{21}$$

The multi-objective function in (21) (i) Assigns OPs priority by allocating the OPs PRBs with high SINRs reflecting their relative priority, (ii) maximizes the sum of the SINRs assigned to all users, and (iii) achieves fairness by assigning healthy users PRBs with comparable SINRs. These objectives were implemented by adding both the summation of a log function of the healthy users' SINRs (i.e. Proportional Fairness) and the weighted sum of the OPs' SINRs (OPs priority).

**Constraints:**

The model satisfies the constraints of the previous model. In addition to equation (17) and the same set of equations for the piecewise linearization that was used before, however, the difference is in the range of users it is applied to (applied only to the normal users).

### 3) Calculating the Received Power

The received signal power (in Watts) $Q_{k,n}^b$ varies according to two elements, namely, the distance between the user and the BS and the channel conditions. The received signal power at the base station is given as

$$Q_{k,n}^b = P\ H_{k,n}^b A_k^b \tag{22}$$

where $H_{k,n}^b$ denotes Rayleigh fading and $A_k^b$ represents power loss due to attenuation (distance-dependent path loss) [64] and is given by (23) and (24), for the MBS and the PBS, respectively, where

$$A\ (dBm) = 128 + 37.6\ \log_{10} \frac{distance(meters)}{1000} \tag{23}$$

$$A\ (dBm) = 140.7 + 36.7\ \log_{10} \frac{distance(meters)}{1000} \tag{24}$$

Equation (25) is used to unify the units by converting the power to Watts.

$$A\ (mw) = 10^{\frac{A(dBm)}{10}} \tag{25}$$

### IV. Results and discussion

We consider a HetNet serving an urban environment, hence the Rayleigh fading channel model with path loss. The results evaluate two scenarios; the first depicts the HetNet state before prioritising the OPs. In this scenario, equal base priority (i.e., weight) of 1 is granted to all users. The second scenario shows the HetNet state after prioritising the OP through the updated priorities according to the value of the tuning factor $\alpha$ and their voted stroke likelihood.

A cloud-based arrangement is assumed where each OP has their personal dataset constructed from their medical history and daily observations over the course of 200 days, with the requirement to periodically extend the dataset by appending recent observations. Moreover, the proposed approach assumes a system that is in operation and the outpatient is being assessed by the voting system where multiple classifiers reside. We divided our dataset into two parts, a training set and a testing set, the training set comprised of 140 entries is used to train, i.e., fit the classifiers, and the test set with 60 entries is used to compare and verify the classifiers' performance. Furthermore, we would like to bring to the reader's attention that the ensemble's role in this work is to report the soft-voted stroke likelihood. Since the outpatients are all under continuous monitoring, they are favoured according to their probability of stroke as long as the system is operational. The OPs' stroke likelihood $P_{voting}$ were 0.42, 0.84, and 0.65 for users 8, 9 and 10 (i.e., OP 1, 2, and 3), respectively. Moreover, the use of equation (12) produced $1.42 \leq UP_k \leq 1.84$, $1.84 \leq UP_k \leq 2.68$, $3.1 \leq UP_k \leq 4.25$, $5.2 \leq UP_k \leq 9.4$ user priorities according to tuning factor values of $\alpha$ of 1, 2, 5, and 10, respectively.

### A. Classifiers Comparison and Evaluation

In this section, we investigate the performance of the methods described in the previous section. There are several performance metrics for ML algorithms and certain metrics are known by more than one name. Since we have a binary classification problem, we refer to a prediction as "positive" if a classifier predicted $P(C = c|F_i = f_i) \geq 0.5$, indicating the occurrence of an *event* (e.g., stroke). Alternatively, if $P(C = c|F_i = f_i) < 0.5$ then the classifier predicted a *no-event* (e.g., no stroke), hence is translated as a "negative" prediction. In order to investigate the classifiers' performance, we use a test dataset of 60 entries where the outcome of all entries (i.e., feature vectors) are known (i.e., observed) to us and register the prediction results. Consequently, there will be four outcomes; (i) a correct positive prediction, named true positive (TP), indicating $P(C|F_i) \geq 0.5$ and an observed output of 1, (ii) an incorrect positive prediction, named false positive (FP), indicating $P(C|F_i) \geq 0.5$ and an observed output of 0, (iii) a correct negative prediction, named true negative (TN), indicating $P(C|F_i) < 0.5$ and an observed output of 0, and (iv) an incorrect negative prediction, named false negative (FN), indicating $P(C|F_i) < 0.5$ and an observed output of 1. The following metrics are computed through the use of these outcomes.

#### 1) Accuracy

Defined as the ratio of true (i.e., correct) predictions to the total number in the dataset and is given in (26). Accuracy measures how well the classifier did in predicting the occurrence of an *event* as well as *no-event*.

$$Accuracy = \frac{TP + TN}{TP + TN + FP + FN} \times 100\% \quad (26)$$

#### 2) Sensitivity, True positive rate (TPR), or recall

Defined as the classifier's ability to pick an *event* of interest. Thus, accurately classifying actual positive values by labelling them as TP (i.e., stroke=1), and is given in (27). In this work, it measures the classifier's ability to correctly classify an individual as *at-risk*.

$$Sensitivity = \frac{TP}{TP + FN} \times 100\% \quad (27)$$

#### 3) Specificity, True negative rate (TNR)

Is a measure of the classifier's ability to pick the occurrence of a *no-event* of interest. In other words, it is the classifier's ability to accurately identify actual negatives (i.e., stroke=0) in the test dataset. Thus, accurately classify an individual as *risk-free*.

$$Specificity = \frac{TN}{TN + FP} \times 100\% \quad (28)$$

#### 4) Precision, Positive-predictive value (PPV)

Answers the question of how many of those who we predicted as at risk are actually at risk? Thus, it is the ratio of accurate positive predictions to the total number of positively-classified feature vectors, as in (29).

$$Precision = \frac{TP}{TP + FP} \times 100\% \quad (29)$$

Precision is a vital measure when the FP's cost is high. In our case, this corresponds to granting a high priority to an outpatient that is not really in a high risk.

#### 5) Negative-predictive value (NPV)

NPV answers the question of how many of those who we predicted as at no risk are actually not at risk. Thus, it is the ratio of feature vectors accurately classified as negative (i.e., TN) to the total number of classifications belonging to class $stroke = 0$, as denoted in (30).

$$NPV = \frac{TN}{TN + FN} \times 100\% \quad (30)$$

#### 6) False-predictive value (FPV)

Also known as false alarm ratio, it represents the rate of misclassifying a class $stroke = 0$ as $stroke = 1$. It measures the frequency of false alarm and is given in (31).

$$FPR = \frac{FP}{FP + TN} \times 100\% \quad (31)$$

#### 7) False-negative rate

A measure telling how erroneous a classifier can be in missing events (i.e., stroke=1). It is the ratio of misclassified positives to the total number of positives, as in (32).

$$FNR = \frac{FN}{FN + TP} \times 100\% \quad (32)$$

#### 8) F1 Score

Defined as a function of both precision and recall values given in (29) and (27), respectively. This score is a measure of the balance between precision and recall as the former highly focuses on true positives, whilst the latter focuses on true negatives. Thus, providing an equal weight for both precision and recall as it is the harmonic mean of the two as given in (33).

$$F1\ Score = \frac{2 \cdot precision \cdot recall}{precision + recall} \quad (33)$$

It should be noted that since there are three separate datasets (one per outpatient), hence, there are not only four classifiers to investigate, but there is also a need to examine the performance of these classifiers over three datasets as illustrated in Table IV.

The proposed SV classifier achieved higher accuracy compared to the other classifiers. Moreover, it had the lowest combined FPR and FNR which motivates its employment in this work. We further scrutinized the accuracy aspect of the proposed SV classifier using 10-folds cross-validation and the results yielded 87.5%, 85.5%, and 88.5%, for the three data sets, respectively.

**TABLE IV.** Comparing the machine learning methods.

| Classifier | Accuracy (%) | Recall (%) | Specificity (%) | Precision (%) | NPV (%) | FPR (%) | FNR (%) | F1 Score |
|---|---|---|---|---|---|---|---|---|
| **OP#1 Training Dataset** | | | | | | | | |
| NB | 82 | 76 | 88.5 | 90 | 74 | 11.5 | 24 | 83 |
| LR | 88 | 85 | 92 | 94 | 82.7 | 7.7 | 15 | 89 |
| DT | 90 | 88 | 92 | 94 | 86 | 7.6 | 11 | 91 |
| SV | **90** | 88 | 92 | 94 | 86 | 7.6 | **11** | **91** |
| **OP#2 Training Dataset** | | | | | | | | |
| NB | 76.7 | 68 | 84.4 | 79 | 75 | 15.6 | 32 | 73 |
| LR | 80 | 75 | 84.4 | 81 | 79 | 15.6 | 25 | 78 |
| DT | 80 | 64 | 94 | 90 | 75 | 6.2 | 36 | 75 |
| SV | **81.7** | 75 | 87.5 | 84 | 80 | 12.5 | **25** | **79** |
| **OP#3 Training Dataset** | | | | | | | | |
| NB | 86.6 | 76 | 94.3 | 90 | 84.6 | 5.7 | 24 | 83 |
| LR | 90 | 88 | 91.4 | 88 | 91.4 | 8.6 | 12 | 88 |
| DT | 91.7 | 84 | 97.1 | 95 | 89.5 | 2.8 | 16 | 89 |
| SV | **93** | 88 | 97.1 | 96 | 92 | 2.8 | **12** | **92** |

B. *Performance Metrics*

While it is important to scrutinize the classifiers at hand and verify their performance, however, given the nature of our work, there are several performance metrics that are more vital than others. Hence we highlight their importance in this section. Accuracy is an important metric in our work due to the fact that it gives a balanced insight on the classifier's overall performance. FNR is the most important metric from the point of view of saving a patient's life, i.e., it tells us the proportion of ill people who are miss-classified. The F1-score takes misclassified entries (i.e., FP and FN) into account. Depending on the application, it can be as important as accuracy in our case.

Before proceeding into the results of the MILP model, it is worth noting that we used the parameters in Table V.

**TABLE V.** MODEL PARAMETERS

| Parameter | Description |
|---|---|
| LTE-A system bandwidth | 1.4 MHz |
| Channel Model | Path Loss [64] and Rayleigh fading [61] |
| No. of MBS | 1 |
| No. of PBS | 2 |
| Number of PRBs per BS | 5 |
| Number of users | 10 |
| Number of normal users ($NU$) | 7 |
| Number of OPs | 3 |
| AWGN ($\sigma_{k,n}^b$) | -162 dBm/Hz [64] |
| The distance between user $k$ and MBS $b$ | (300 - 600) m |
| The distance between user $k$ and PBS $b$ | (40-100)m |
| Maximum transmission power per connection $PM$ | 23 dBm [64] |
| UE transmission power per PRB | 17 dBm |
| Minimum SINR for the reliability aware PF approach ($\psi$) | 127 |
| Base (i.e. normal user priority) weight | 1 |
| Outpatient priority $UP_k$ calculation method | Soft Voting Classifier |
| OP observation period | 200 Days |
| $\alpha$ values | 1, 2, 5, and 10 |

C. *The WSRMax Approach*

*1) Before Prioritizing the OPs*

This scenario considers the operation of a HetNet where all users share the same base user weight (i.e. priority) of 1. The results in Fig. 6. indicate that the OPs (represented by users 8, 9, and 10) are assigned PRBs of comparable gains resulting in near-average SINRs. This is due to the fact that the MILP's aim is to maximize the HetNet's overall SINR. In order to measure fairness, we considered emphasizing the Standard Deviation (SD) of the users' SINRs, hence, to quantify how close the calculated SINR values are to the mean, in this case, the SD was 195. Moreover, an extensive sensitivity analysis was carried out for the 300 independent realizations of the channel and the results with 95% confidence intervals per user are indicated in Fig. 6. The average SINR lied between 2166 and 2691.

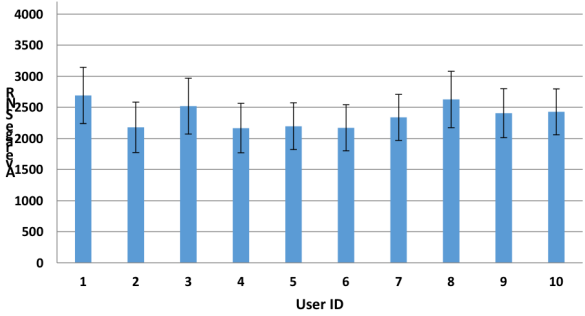

**FIGURE 5. User SINR before OP Prioritization (WSRMax Approach)**

#### 2) After Prioritizing the OPs

The goal in this scenario is to utilize BDA/ML to prioritize the OPs over normal users by means of the ensemble system. As a result, high gain PRBs will be allocated to the OPs according to their risk factor guaranteeing them high-level SINRs. Comparing Fig. 6. and Fig. 7. clearly highlights that the OPs (i.e., users 8, 9, and 10) were granted PRBs with high SINRs. The overall system performance is a trade-off (optimally-selected) between guaranteeing the assignment of high SINRs to the OPs versus decrease in the average SINR (between 2% ($\alpha = 1$) and 19% ($\alpha = 10$) in comparison to the SINR in the first scenario.

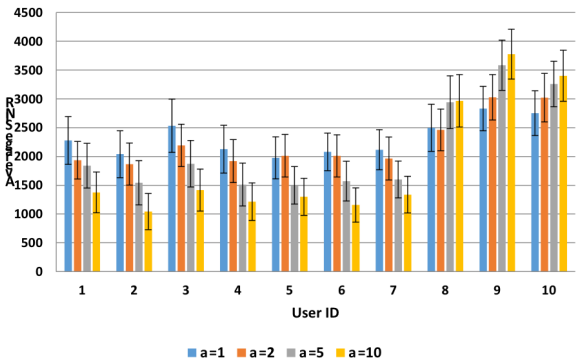

**FIGURE 6. User SINR after OP Prioritization (WSRMax Approach)**

The reason behind the reduction in the average SINR is because the system was forced to a PRB assignment scheme where the maximization of the OPs' individual SINRs is prioritized over the total SINR. Fairness between normal users was marginally impacted in this approach as will be shown in the following subsection. The impact of converting the probability of stroke to a risk factor and using several values of the tuning factor (i.e. $\alpha = 1, 2, 5$ and $10$) can be observed by comparing the increase in the OPs' average SINRs. Taking the case of user 9 (the most critical user with a probability of 0.84) having an SINR lower than users 1, 3, 8, and 10, the average SINR witnessed an increase from 17% ($\alpha = 1$) to 57% ($\alpha = 10$) granting this user an average SINR higher than all users. Individual users had an average SINR ranging from 1042 to 3776 for $\alpha = 10$.

#### 3) The Impact of α on Fairness and SINR

The parameter $\alpha$ is a tuning factor that is used to convert the minute value of the voted probability (i.e., $P_{voting}$) of stroke acquired from the ensemble system to a risk factor as depicted in equation (12). Moreover, this parameter enables the reciprocity between the average SINR and the attainable fairness among the users quantified by the SD. We used different values of $\alpha$ to study the effects on the SD and the average SINR. We examined the effects of using different vales of $\alpha$ on the SD and the average SINR as shown in Fig. 8. and in Fig. 9.

Increasing the value of $\alpha$ forced the system to concentrate on the OPs. Accordingly, the system's overall SINR was optimally traded-off to increase the OPs' SINRs while minimally impacting fairness among users as shown in Fig. 8. It should be noted that examining the OPs' SINRs and comparing them against their corresponding risk factor values reveals an increase in the SINR in an order conforming to that depicted in Fig. 9, where the PRB assignment granting the highest SINR was allocated to user 9 which is the user with the highest risk factor (priority). Furthermore, user 8 which has the lowest risk factor among the three OPs was given the lowest SINR among the OPs and very close to the system's average SINR. As the value of $\alpha$ increased (i.e., $\alpha = 5, 10$), user 8 is granted higher SINRs in comparison to other healthy users.

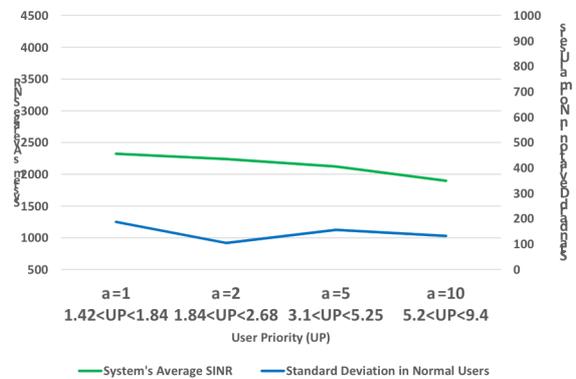

**FIGURE 7. Effects of changing α on average SINR and fairness (WSRMax Approach)**

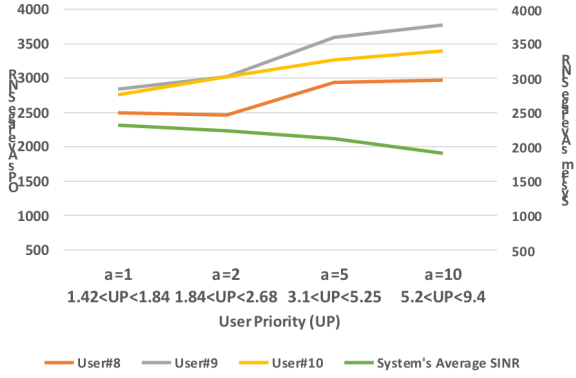

**FIGURE 8.** The impact of α, both user and average SINR (WSRMax Approach)

### D. The PF Approach

#### 1) Before Prioritizing the OPs

In this scenario, the goal is to maximize the logarithmic sum of the user's SINRs. Thus, no priority is given to any user in particular. Fairness is applied as a consequence due to the nature of the natural log in the objective function in (19). The results depicted in Fig. 10. are in agreement with the ones depicted in Fig. 6.

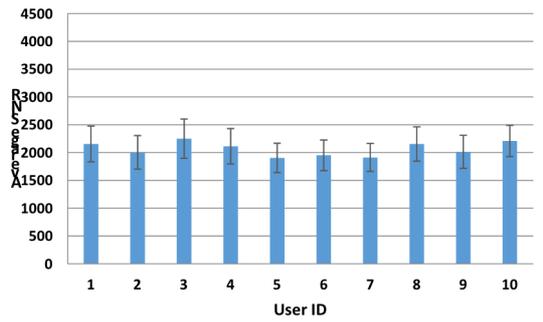

**FIGURE 9.** User SINR before OP Prioritization (PF Approach)

However, a 46% reduction in the SD is observed when comparing this scenario and the one in Subsection IV.B.1. The average SINR ranged between 1905 and 2251. Sensitivity analysis was implemented over 300 different realizations of the HetNet. The results with a 95% confidence interval are illustrated in Fig. 10.

#### 2) After Prioritizing the OPs

In this approach, the OPs are prioritized according to their risk factors using the objective function in (21). Therefore, the OPs are granted high-gain PRBs resulting in high SINRs as illustrated in Fig.11. The OPs' SINRs were boosted by up to 95% observed by user 9 with $\alpha = 10$. However, the average system SINR ranged between 1093 ($\alpha = 1$) and 1113 ($\alpha = 10$).

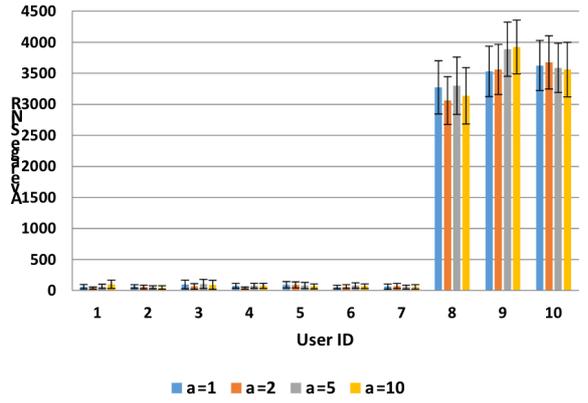

**FIGURE 10.** User SINR after OP Prioritization in linear Scale (PF Approach)

The healthy users were noticeably affected by the intrinsic nature of the natural log, and the exclusion of the OPs from the logarithmic term in the objective function resulted in granting the healthy users lower SINRs in comparison to the OPs' SINRs. Fig. 12. depicts the average users' SINR in a logarithmic scale where narrower confidence intervals can be observed in this approach.

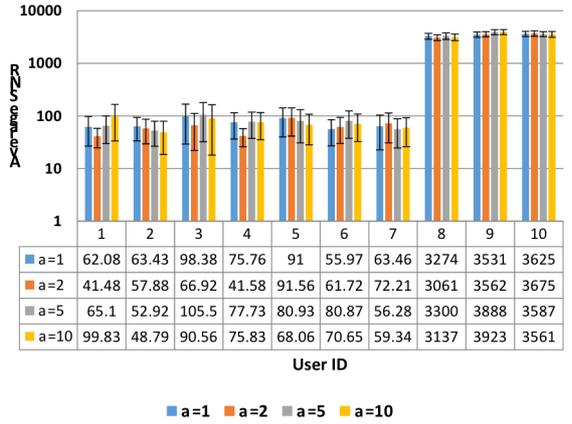

**FIGURE 11.** User SINR after OP Prioritization in logarithmic Scale (PF Approach)

#### 3) The Impact of α on Fairness and SINR

Increasing the OPs' priority by adjusting the tuning factor $\alpha$ has similar effects to the ones observed in the previous subsection. Using the PF approach, boosts the OPs' SINRs by up to 95%, but has resulted in reducing the overall system SINR by up to 48% while maintaining a good fairness interpreted as a stable and very low SD as illustrated in Fig.13. Observing Fig. 14, it can be clearly seen that the OPs' are granted SINRs approximately three times the system's average SINR. Furthermore, the analogy between the priorities (weights) granted to the OPs and the corresponding increase in their SINRs is highlighted.

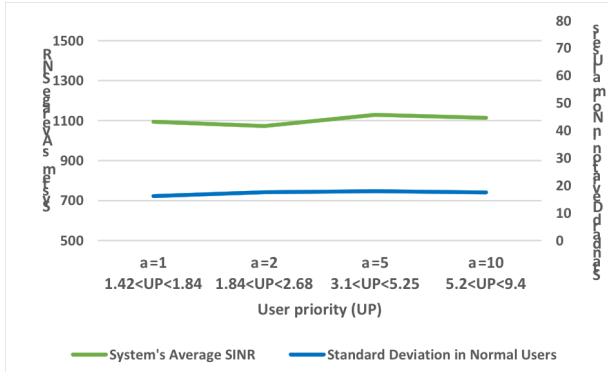

**FIGURE 12. Effects of changing α on average SINR and fairness (PF Approach)**

It should be noted that user 9, despite having a higher priority than user 10, was assigned a PRB with SINR very close to the SINR of user 10 when $\alpha = 1, 2$. This is due to the fact that user 10 has already better channel conditions than user 9 as indicated in Fig. 10. Thus, it would require higher values of the tuning factor $\alpha$ to bias the system towards user 9 and this can be seen in $\alpha = 5$ and 10 in Fig. 14.

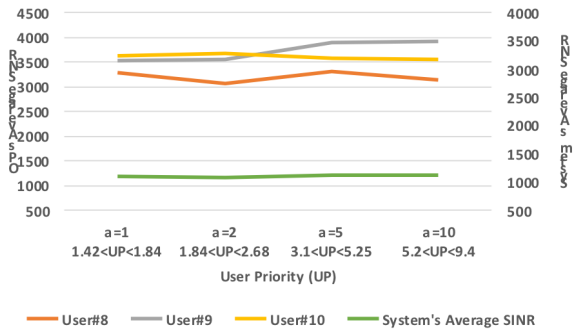

**FIGURE 13. The impact of α, both user and average SINR (PF Approach)**

### E. The Reliability-aware PF Approach

In this approach, we are enhancing the SINR values for the normal users that are impacted by the logarithmic sum. This is done by setting a minimum SINR where the users that are subjected to this constraint have guaranteed reliable service levels [65].

#### 1) Before Prioritizing the OPs

This approach shares the same objective of the PF approach in Section IV.D.1. However, a constraint is added to the model guaranteeing a minimum SINR of 21 dB for all users. The results depicted in Fig. 15 show a similar trend to the ones illustrated in Fig. 10. However, preserving a minimum SINR level with no prioritization means that there is a slight impact on the system-wide SINR. Thusly, we observed a 5% decrease in the system's average SINR for the PF approach before and after introducing reliability. Here

$$S_k \geq \psi .$$
$$\forall\, k \in \mathcal{K}: 1 \leq k \leq NU \tag{34}$$

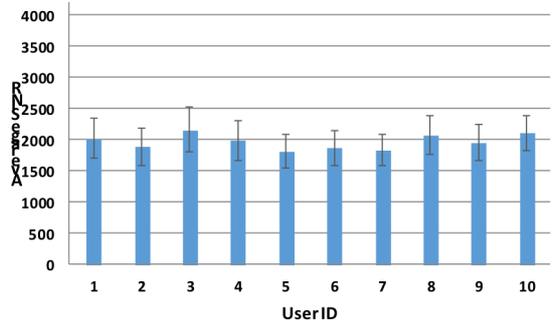

**FIGURE 14. User SINR before OP Prioritization (Reliability-aware PF Approach)**

#### 2) After Prioritizing the OPs

The impact of the natural logarithm on healthy users motivated the inclusion of a constraint guaranteeing the minimum SINR level as in [65]. This results in an additional level of reliability with fairness in the PF approach, and here

$$S_k \geq \psi .$$
$$\forall\, k \in \mathcal{K}: 1 \leq k \leq NU \tag{35}$$

Constraint (35) works under the objective in (21) to guarantee a minimum SINR level specified by the parameter $\psi$. The result of introducing this constraint is shown in Fig. 16.

The OPs' SINRs are boosted by up to 23% observed by user 9 with $\alpha = 10$. However, the OPs' SINRs are now reduced in comparison with the previous scenario before introducing reliability as shown in Fig. 11. The results show narrower confidence intervals than under the WSRMax approach indicating a further reduction in the error values.

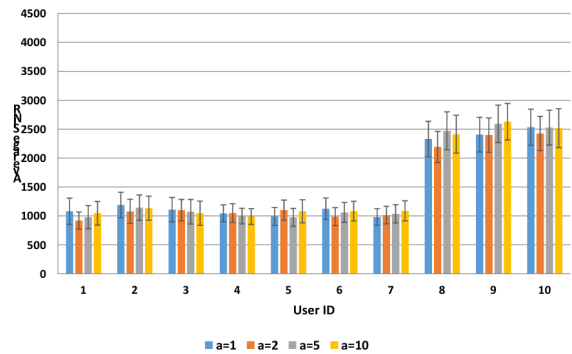

**FIGURE 15. User SINR after OP Prioritization (Reliability-aware PF Approach)**

### 3) The Impact of α on Fairness and SINR

Introducing the reliability aspect to the PF approach resulted in improving the system's average SINR with a marginal increase in the SD. However, better fairness is observed when increasing the tuning factor α as indicated in Fig. 17. Furthermore, the average SINR is increased by 32% in comparison to the reliability-unaware PF approach. The OPs' SINRs witnessed a 30% increase when employing the reliability-aware PG approach as shown in Fig. 18. Moreover, the OPs were granted SINRs that are approximately 70% higher than the system's average SINR

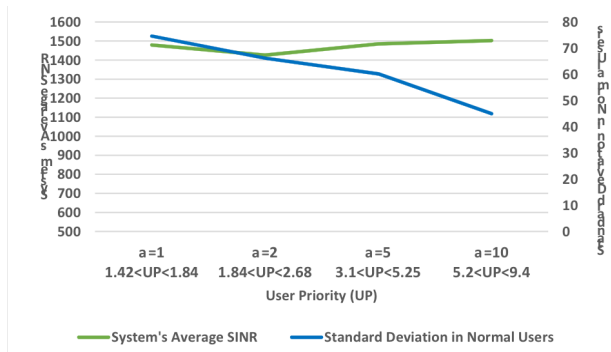

**FIGURE 16. Effects of changing α on average SINR and fairness (Reliability-aware PF Approach)**

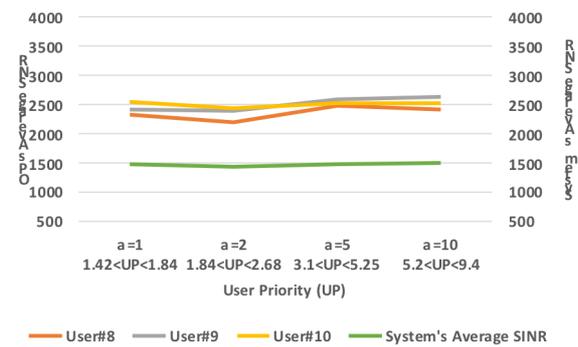

**FIGURE 17. The impact of α, both user and average SINR (Reliability-aware PF Approach)**

## V. CONCLUSIONS

This work introduced two interdisciplinary approaches to transform conventional HetNets in 6G by endowing them with a user-centric machine learning dimension. To that end, a BDA-powered framework was proposed to play part in uplink radio resource allocation optimization of HetNets. The target is to prioritize stroke outpatients within the HetNet to provide them with the optimal wireless resources. Moreover, the assigned resources should be proportional to the severity of the patients' medical state (i.e., stroke likelihood), which is predicted using an ensemble system classifying readings of vital signs acquired from body-attached and nearby IoT sensors. Two approaches, namely, the WSRMax and the PF are presented and compared in terms of fairness and in terms of the average SINR (both at the system and the user level).

The WSRMax approach enhanced the OP's average SINR by up to 57%, whereas the PF approach improved the SINR by up to 95%. Depending on the value of tuning factor $\alpha$, normal users reported an average SINR ranging between 2163 and 1263 using the WSRMax approach, while the reliability-aware PF approach attained an SINR ranging from 1089 to 1066 (depending on $\alpha$). Using the SD to quantify fairness among users, the WSRMax scored between 104 and 156, while the reliability-aware PF approach ranged between 44 and 74. Furthermore, to add confidence in the estimated probability of stroke, the ensemble system is examined and the voting classifier yielded up to 93% accuracy, a false positive rate of 2.8% and a false negative rate of 11%.

### ACKNOWLEDGMENTS

The authors would like to thank EPSRC for partly funding this work. All data are provided in full in the results section of this paper.